# Action for perception : influence of handedness in visuo-auditory sensory substitution

*Sylvain Hanneton[1] & Claudia Munoz[1]*
[1]Laboratoire de Neurophysique et Physiologie. CNRS UMR 8119 et Université René Descartes. UFR Biomédicale des Saints Pères, 45 rue des Saints-Pères. 75270 Paris, France
sylvain.hanneton@univ-paris5.fr

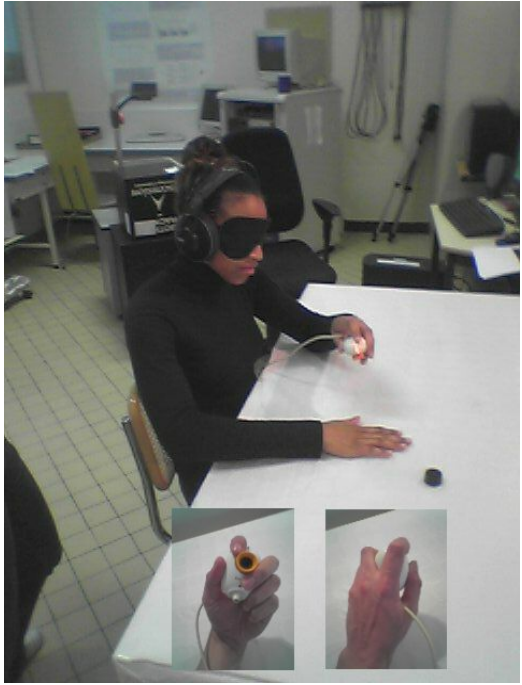

**Figure 1. Experimental setup and the imposed hand grasping posture for holding the webcam.**

In this preliminary study we address the question of the influence of handedness on the localization of targets perceived through a visuo-auditory substitution device. Participants hold the device in one hand in order to explore the environment and to perceive the target. They point to the estimated location of the target with the other hand. This location can be considered as an enactive knowledge since it is gained through perception-action interactions with the environment. The handedness can influence the accuracy of the pointing : has the device to be hold in the right or left hand? There are two possible main results. (1) Participants are more accurate with the device in the left hand because *pointing movements* are more skillful with the dominant hand. (2) Participants are more accurate with the device in the right hand because *exploratory movements* (perceptive movements) are more precisely controlled with the right hand. Enaction theory assumes that action for perception is crucial to establish an enactive knowledge. According to this theory, the dominant hand has to be used for a fine control of perceptive movements rather than for pointing movements. Consequently we expect to obtain the second result. In an other context, right handed rifle shooters with a dominant left eye were shown to be more accurate if they hold the rifle in the left arm (Porac & Coren, 1981).

**Methods**

The VIBE device converts a video stream to a stereophonic sound stream. It requires only a standard webcam and a standard computer (see Auvray et al. 2005 for details). Results presented here concerned three young right-handed females. Participants were instructed to point to targets perceived and memorized either visually ("vision" experimental condition, VEC) or via the VIBE device ("prosthesis" condition, PEC). In the VEC condition, subjects were asked to observe the target during three seconds, to close the eyes and to point immediately to the estimated position of the target with either the left or right index. In the PEC condition, participants were blindfolded, wore closed headphones, and held the webcam in the right or left hand with an imposed grasping posture (figure 1). The elbow of the arm that hold the webcam had to keep at a specific location on the table. Participants had 15 seconds to explore the environment and then pointed to the estimated target location. Each participant did 45 trials (3 target positions x 15 repetitions) in four experimental conditions (VEC or PEC and pointing with the left or the right hand). We studied the influence of experimental conditions on distance to target, on constant and variables pointing errors and on confidence ellipse for each target and each experimental condition. Considering confidence ellipses (a linear local estimates of errors distributions) can give an access to perception distortion induced by "defaults" of perceptive organs, memory or actuators and to the structure of the spatial representation of targets (McIntyre et al., 1998).

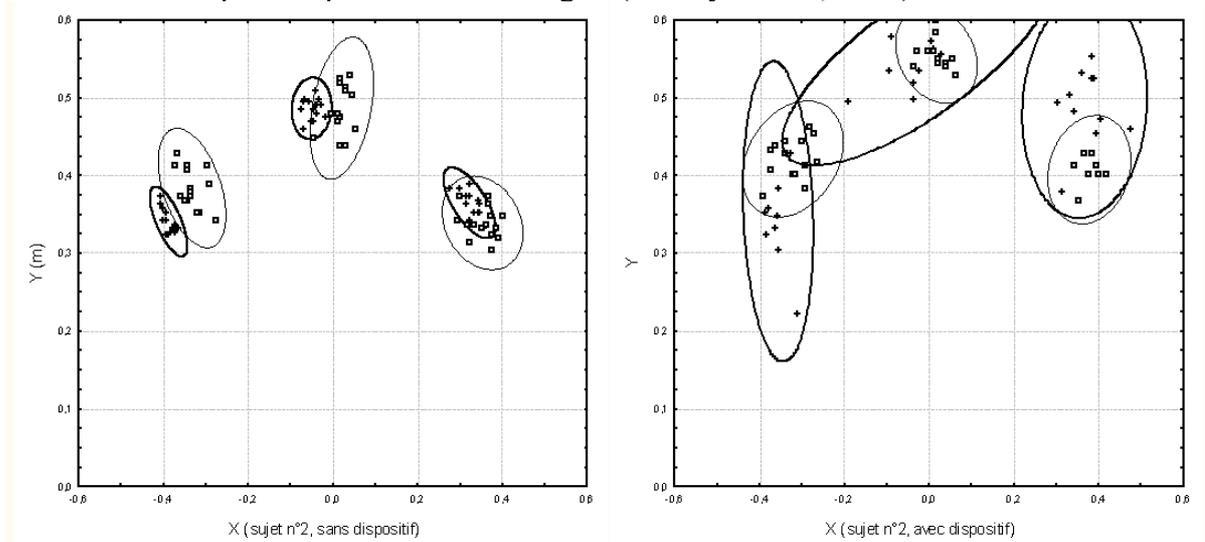

**Figure 2. Pointing confidence ellipses obtained for the second participant when pointing either to visually remembered targets (VEC, left) or to targets perceived trough the sensory substitution device (PEC, right). Bold ellipses : pointing with the right hand. Solid ellipses : pointing with the left hand.**

**Results**

The mean distance to target is obviously lower in the VEC condition than in the PEC condition (table 1). In the PEC condition, the mean distance to target is lower when pointing with the left hand (webcam in the right hand) than in the opposite condition. You can also notice that the variance of the distance to target is systematically lower in this condition. Characteristics of the confidence ellipses are very variable across subjects, targets and experimental conditions. However we have to emphasize that in the PEC condition, participants were more precise in seven cases over nine (3 targets x 3 participants) when holding the webcam in the right hand. The "inversion" of the handedness is particularly clear for the participant #2 (figure 2).

**Table 1. Means and standard deviations for the different error measures.**

|  | Condition VEC (vision) | | Condition PEC (prosthesis) | |
| --- | --- | --- | --- | --- |
| Pointing with → | Right hand | Left hand | Right hand | Left hand |
| Distance to target (cm) | 3.87 ± 1.81 | 3.99 ± 2.17 | 6.48 ± 5.00 | 6.06 ± 3.1 |
| Absolute distance error (cm) | 1.86 ± 1.41 | 2.41 ± 1.78 | 3.83 ± 3.88 | 4.06 ± 2.7 |
| Absolute direction error (deg) | 3.51 ± 2.25 | 3.25 ± 2.35 | 5.07 ± 4.47 | 4.22 ± 3.1 |

**Discussion**

This preliminary results support our hypothesis that pointing is more accurate when the device is held in the right dominant hand. Dexterity has to be attributed to the active part of the perceptive system. This study has obviously to be completed but it shows how the concept of enaction is important and how it can be experimentaly addressed in the field of sensory substitution.